\documentclass[twocolumn]{aastex631}

\usepackage{soul}

\begin{document}

\title{The weakness of soft X-ray intensity: possible physical reason for weak line quasars}

\author[0000-0002-2581-8154]{Jiancheng Wu}
\affiliation{Department of Astronomy, School of Physics, Huazhong University of Science and Technology, Luoyu Road 1037, Wuhan, China}

\author[0000-0003-4773-4987]{Qingwen Wu$^*$}
\affiliation{Department of Astronomy, School of Physics, Huazhong University of Science and Technology, Luoyu Road 1037, Wuhan, China}

\author[0000-0002-2006-1615]{Chichuan Jin}
\affiliation{National Astronomical Observatories, Chinese Academy of Sciences, 20A Datun Road, Beijing 100101, China}
\affiliation{School of Astronomy and Space Sciences, University of Chinese Academy of Sciences, 19A Yuquan Road, Beijing 100049, China}

\author[0000-0001-7349-4695]{Jianfeng Wu}
\affiliation{Department of Astronomy, Xiamen University, Xiamen, Fujian 361005, People's Republic of China}

\author[0000-0003-3440-1526]{Weihua Lei}
\affiliation{Department of Astronomy, School of Physics, Huazhong University of Science and Technology, Luoyu Road 1037, Wuhan, China}

\author[0000-0002-2355-3498]{Xinwu Cao}
\affiliation{Institute for Astronomy, School of Physics, Zhejiang University, 866 Yuhangtang Road, Hangzhou 310058, People’s Republic of China}

\author[0000-0001-7350-8380]{Xiao Fan}
\affiliation{Department of Astronomy, School of Physics, Huazhong University of Science and Technology, Luoyu Road 1037, Wuhan, China}

\author[0009-0003-0516-5074]{Xiangli Lei}
\affiliation{Department of Astronomy, School of Physics, Huazhong University of Science and Technology, Luoyu Road 1037, Wuhan, China}

\author[0000-0001-5019-4729]{Mengye Wang}
\affiliation{Department of Astronomy, School of Physics, Huazhong University of Science and Technology, Luoyu Road 1037, Wuhan, China}

\author[0000-0002-8064-0547]{Hanrui Xue}
\affiliation{Department of Astronomy, School of Physics, Huazhong University of Science and Technology, Luoyu Road 1037, Wuhan, China}

\author[0000-0001-8879-368X]{Bing Lyu}
\affiliation{Kavli Institute for Astronomy and Astrophysics, Peking University, Beijing 100871, Peoples Republic of China}

\begin{abstract}

Weak-line quasars (WLQs) are a notable group of active galactic nuclei (AGNs) that show unusually weak UV lines even though their optical-UV continuum shapes are similar to those of typical quasars. The physical mechanism for WLQs is an unsolved puzzle in the  AGN unified model.  We explore the properties of UV emission lines by performing extensive photoionization calculations based on Cloudy simulation with different spectral energy distributions (SEDs) of AGNs. The AGN continua are built from several observational empirical correlations, where the black-body emission from the cold disk, the power-law emission from the hot corona, and a soft X-ray excess component are considered. We find that the equivalent width (EW) of C {\footnotesize IV} from our models is systematically lower than observational values if the component of soft X-ray excess is neglected. The EW will increase several times and is roughly consistent with the observations after considering the soft X-ray excess component as constrained from normal type I AGNs. We find that the UV lines are weak for QSOs with quite large BH mass (e.g., $M_{\rm BH}>10^9M_{\odot}$) and weak soft X-ray emission due to the deficit of ionizing photons. As an example, we present the strength of C {\footnotesize IV} based on the multi-band SEDs for three nearby weak-line AGNs, where the weaker soft X-ray emission normally predicts the weaker lines.

\end{abstract}

\keywords{Active galactic nuclei (16), Supermassive black holes (1663), Quasars (1319), X-ray active galactic nuclei (2035), Accretion (14), Line intensities (2084)}

\section{Introduction} \label{sec:intro}

Active Galactic Nuclei (AGNs) are the most luminous persistent sources of electromagnetic radiation in the Universe, which is believed to be powered by accretion onto supermassive black holes \citep[SMBHs, e.g.,][]{Shakura1973}. One of most important features of AGNs is that their optical spectra exhibit fruitful broad and narrow emission lines. The type I AGNs show both broad lines  (i.e., Full Width at Half Maximum (FWHM)$\rm {\gtrsim 1000 km \, s^{-1}}$) and narrow lines and type II AGNs show only narrow lines, which are unified based on orientation effect \citep{Antonucci1993, Urry1995}. The Sloan Digital Sky Survey (SDSS) project has discovered millions of quasars in the past few decades. \cite{Aleksandar2009} studied the equivalent widths (EW) of UV emission lines for a sample quasars and found that the average equivalent widths are $63.6\text{\AA}$ and  $41.9\text{\AA}$ for Ly$\alpha$+N {\footnotesize V} and C {\footnotesize IV} respectively. However, a small fraction of quasars show extremely weak ultraviolet high-ionization lines, where EW(C {\footnotesize IV}) $<10\text{\AA}$ and ${\rm EW(Ly\alpha}$+N {\footnotesize V})$<15.4\text{\AA}$ even their properties of optical continuum and H$\alpha$ lines are similar to normal quasars \citep[e.g.,][]{Aleksandar2009, Liu2013}. These sources are 3$\sigma$ outliers at the low-end of log-normal EW distribution \citep[e.g.,][]{Aleksandar2009}, which are called as weak line quasars \citep[WLQs, e.g.,][]{Fan1999, Anderson2001, Collinge2005, Shemmer2010, Wu2011, Wu2012, Luo2015}. Most of WLQs lie at high-redshifts \citep[e.g.,][]{Fan1999, Aleksandar2009}. However,  more and more intermediate- and low-redshift WLQ are also found \citep[e.g.,][]{Plotkin2010, Plotkin2010_2, Paul2022}, even though the fraction of WLQs seems to increase with increasing redshift \citep{Aleksandar2009,Wu2012}.

The physical mechanism for the weak or absent emission lines is still quite unclear so far. One possible physical reason is the underdeveloped broad line region in these AGNs with a lower covering factor of BLR clouds, where WLQs may belong to an early stage in AGN evolution and the BLR was not yet fully formed \citep[e.g.,][]{Shemmer2010, Hryniewicz2010, Zhang2016, Kumar2023}. If this is the case, most emission lines, including the low-ionization lines, in WLQs should be weaker than those of normal quasars. Some WLQs indeed show weaker H$\beta$ lines \citep{Shemmer2010, Plotkin2015}. However, many WLQs show quite strong H$\alpha$ and H$\beta$ lines \citep{McDowell1995, Liu2013}, which suggests that the underdeveloped BLR should be not the dominant or the only reason for all WLQs. The second possible reason is the inefficient photo-ionization of photons. \cite{Wu2011} and \cite{Luo2015} proposed that the ``shielding" gas exists between the X-ray corona and BLR, which may be related to the inner geometrically thick super-Eddington slim disk \citep[e.g.,][]{Abramowicz1988, Czerny2019, Feng2019} and/or the wide-angle optically thick winds \citep[e.g.,][]{Murray1995, Giustini2019, Jiang2019}. The ionization photons from the inner disk/corona will be absorbed by the ``shielding" gas \citep{Wu2011, Luo2015, Ni2018}. The slim disk will be geometrical thick for the case of super Eddington accretion (e.g., $H/R\sim0.5$ for $\dot{M}\sim5\dot{M}_{\rm Edd}$). The more careful estimations for the extinction from the possible ``shielding" gas, the mass of SMBHs and bolometric luminosities will help to confirm whether these WLQs stay in a slim state or not. The absence of UV ionization photons can also be caused by the colder accretion disk surrounding very massive non-spinning SMBHs \citep[][]{Laor2011}. The weaker hard X-ray emission may be also one of the reasons for a deficit of ionization photons. About half of the WLQs have evident lower X-ray luminosities compared to the expectation from the $L_{\nu}(2500{\text \AA}) - \alpha_{\rm ox}$ relation, where $\alpha_{\rm ox}=-0.3838\,{\rm log}(F_{\rm 2500\text{\AA}}/F_{2 {\rm keV}})$ \citep{Luo2015, Ni2018, Ni2022}. The weaker X-ray can either be intrinsic or caused by the shielding gas \citep[e.g.,][]{Luo2015}, where the hard X-ray observations (e.g., NuSTAR) can discriminate them. It should be noted that the WLQs are not the same with beamed BL Lacs, where they have intrinsically weak emission lines or their emission lines are diluted by the emission from the relativistic beaming jets \citep[e.g., ][]{Londish2004, Shemmer2006, Lane2011, Wu2012, Kumar2016}. \cite{Kumar2016} explored the intranight optical variability and polarization property for a couple of radio quiet WLQs, which are different from the typical BL Lacs \citep[see also, ][]{Gopal2013, Chand2014, Kumar2015}. Therefore, the WLQs should be not caused by the diluted emission from the relativistic beaming jets.


The SEDs of UV bump and power-law component have been well built for many AGNs, which dominantly come from cold disk and hot corona respectively. It should be noted that the soft X-ray excess below 2 keV concerning the extrapolation of the hard X-ray spectrum is also a very common feature among type I AGNs even though its radiative mechanism is unclear. It should be noted that this component is always neglected in exploring the emission lines in the photon ionization model even though we know that the high-ionization UV lines are very sensitive to the spectrum from extreme UV to soft X-ray bands. We aim to explore the properties of UV emission lines by performing extensive photoionization calculations based on SEDs and covering factors as constrained from several observational empirical correlations, where we also consider the soft X-ray excess as that constrained from X-ray observations for type I AGNs. This paper is structured as follows: Sec. \ref{sec:model} presents the model of the incident AGN SEDs and the photon ionization model for broad emission lines. Sec. \ref{sec:result} describes our simulation results. Conclusion and discussion are presented in Sec. \ref{sec:discussion}.

\begin{figure*}[htb]
\includegraphics[scale=0.6]{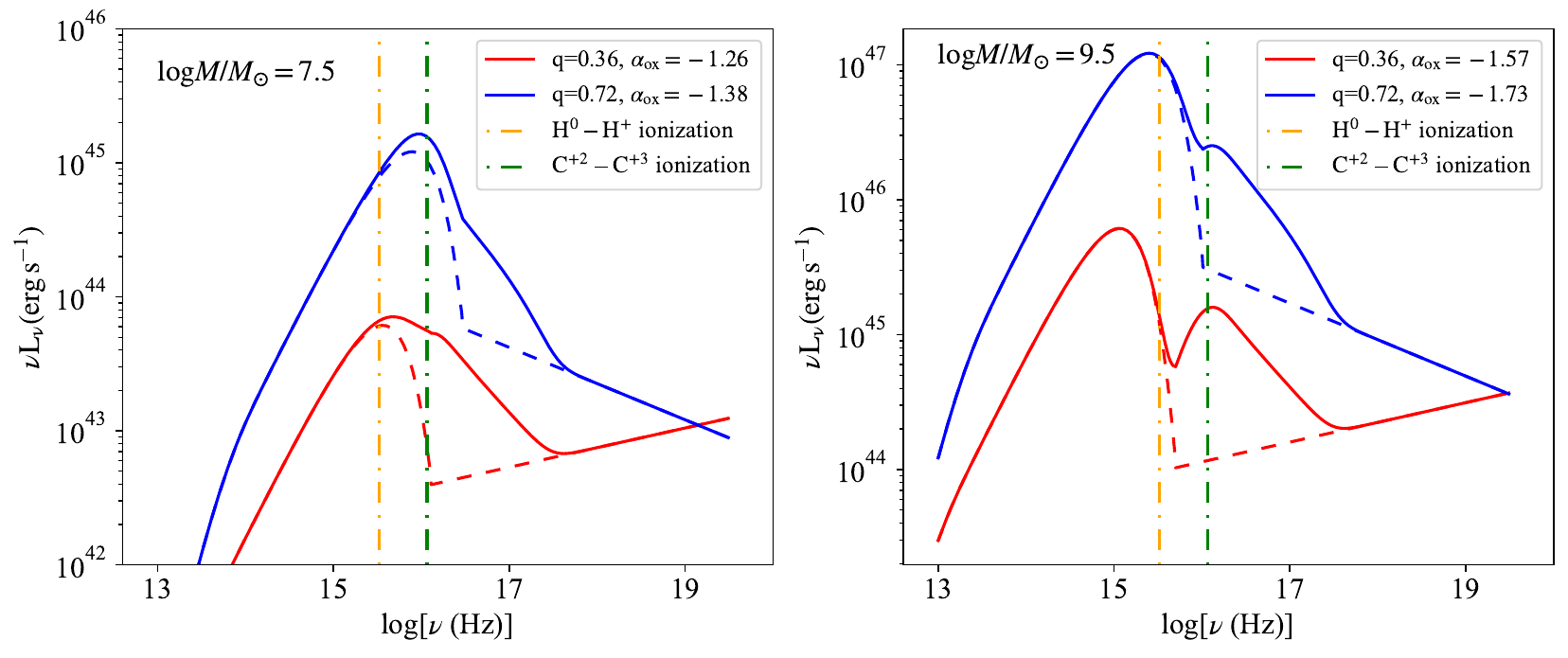}
\caption{The typical SEDs with different parameters are adopted in Cloudy simulations. The blue and red lines represent the SEDs for $\dot{m}=1$ and  $\dot{m}=0.05$ respectively with $10^{7.5}M_{\odot}$ (left panel) and $10^{9.5}M_{\odot}$(right panel), respectively. In both panels, the dashed line represents the SEDs without soft X-ray excess (i.e., $q=0$). The orange and green dot-dashed represent the ionization energy for H and C {\footnotesize IV}, respectively.
\label{sed}}
\end{figure*}

\section{Photoionization Model} \label{sec:model}

The photoionization code is widely adopted to investigate the optical emission lines in AGNs \citep{Ferland1998}. In this work, we generate various SEDs as the incident continuum for Cloudy simulation, where the incident SEDs and photoionization model will be described in more detail as follows.

\subsection{Incident AGN continuum}
The optical-UV emission of bright AGNs is believed mainly from the cold optically thick accretion disk, where the thermal emission is the origin for the so-called  ``big blue bump". We calculate the optical-UV spectrum based on slim disk models, which can cover a wide range of accretion rates from sub-Eddington to super-Eddington accretion\citep[e.g.,][]{Abramowicz1988}. We calculate the hard X-ray spectrum based on two observational correlations, where the 2 keV emission is calculated from $\alpha_{\rm ox} = -0.14\times {\rm log}(L(2500{\text \AA})) + 2.64$ \citep{Steffen2006} and the spectral index is derived from $\Gamma_{2-10{\rm keV}} = 0.32 \times \lambda_{\rm Edd} + 2.27$ from \cite{Brightman2013}, where $\lambda_{\rm Edd}$ is the bolometric Eddington ratio. The X-ray spectra of most AGNs also exhibit a ``soft excess" at energies below $\sim$2 keV, which sits above what would be expected from the hard X-ray power-law. To mimic this soft X-ray component, we simply adopt a cool, optically thick Comptonization component \citep[COMPTT model in XSPEC;][]{Titarchuk1994}, where the input soft photon temperature tied to the blackbody is 0.01 keV, the plasma temperature is 0.2 keV and the plasma optical depth is 15 \citep[e.g.,][]{Chris2012, Jin2022}. The observations suggest that the intensity of soft X-ray excess is normally correlated with the bolometric Eddington ratios \citep{Boissay2016}. We set the intensity ratio of soft excess to power-law component with a parameter of $q = F^{\rm SE}/F^{\rm PL}$ in 0.5--2 keV bands, where we also use the empirical correlation of $q = 0.28 \times {\rm log}\lambda_{\rm Edd} + 0.72$ as constrained from the observations on type I AGNs \citep[][]{Boissay2016}.

In building the incident AGN SEDs, we adopt uniform random distributions of BH mass from $10^7-10^{10}M_{\odot}$ and dimensionless Eddington-scaled accretion rate $\dot{m}=0.02-2$ in log space. The broader parameter space will help to generate various ionization SEDs for incident continuum in Cloudy simulation, which will help us to understand the dependence of the intensities of emission lines on the model parameters. For parameters $\alpha_{\rm ox}$ and $q$, we adopt above observed empirical correlations with standard scatters. For statistical study, we generate 200 incident AGN SEDs for Cloudy simulations, where the distribution of the parameters is given in Fig. \ref{para_dis}. In Fig. \ref{sed}, we present several SEDs for the accretion rate of $\dot{m}=0.05$ and 1 with BH mass of $10^{7.5}M_{\odot}$ and $10^{9.5}M_{\odot}$ respectively, where the parameters of $q$, $\alpha_{\rm ox}$ are derived from above observational empirical correlations. 


\subsection{BLR model and broad-line simulation}
\begin{figure*}[htb]
\includegraphics[scale=0.62]{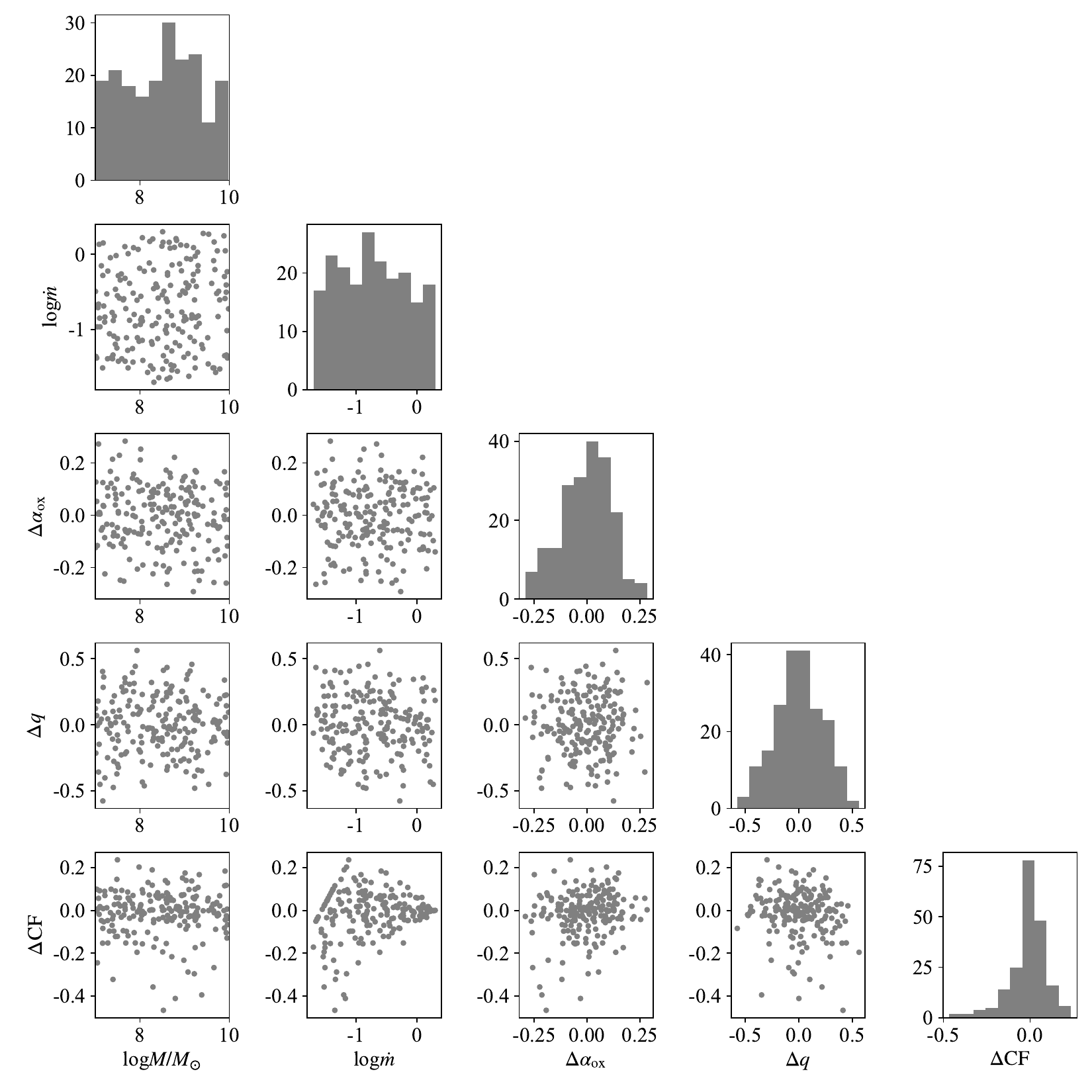}
\caption{The parameter distributions are adopted in our simulations, where $\Delta \alpha_{\rm ox}$, $\Delta q$ and $\Delta {\rm CF}$ are the deviations from the adopted empirical correlations. 
\label{para_dis}}
\end{figure*}

The photoionization models are calculated using version 23.00 of Cloudy, which is described in \cite{Chatzikos2023}. The Locally Optimally Emitting Cloud (LOC) was adopted in calculating the line intensities, where for any given line significant emission occurs only in a narrow range of photoionization parameters \citep[the optimally emitting clouds,][]{Baldwin1995, Ferguson1997, Korista2000, Ferland2003}. The total line luminosity emitted by the BLR is 
\begin{equation} \label{equ2}
    L_{\rm line} \propto \int{\int{r^2 F(r,n) f(r) g(n) dn dr}},
\end{equation} 
\citep[see][for more details]{Bottorff2002}. $F(r, n)$ is the line flux from a single cloud surface with hydrogen density $n$ and distance from the continuum source $r$. The radial differential covering factor is $f(r) \propto r^{\Gamma}$, and the fraction of the clouds with density $n$ is $g(n) \propto n^{\beta}$ (density covering factor). In this work, we adopt the typical values of $\Gamma=-1$ and $\beta=-1$ as suggested in \cite{Baldwin1995} and \cite{Korista2000}. It should be noted that the line intensities are also affected by the BLR geometry or covering factor, where the cloud covering factor normally decreases as the Eddington ratio increases as constrained from observations \citep{Ferland2020, Ezhikode2017}. In this work, the covering factor is adopted to be ${\rm CF} = -0.53\,{\rm log} \dot{m} + 0.17$, which is constrained from the H$\beta$ lines in \cite{Ferland2020}.  In our simulations, we also introduce a Gaussian random factor for the CF with upper and lower limits of CF = 1 and 0.001, respectively (see Fig. \ref{para_dis}).

For BLR size, we adopt $r_{\rm in} = 0.1r_{\rm BLR}$ and $r_{\rm out} = 10r_{\rm BLR}$, where $r_{\rm BLR}$ is calculated from the reverberation mapping methods ${\rm log}(r_{\rm BLR}/{\rm lt-days}) = 1.53 + 0.53\times{\rm log}(L_{5100}/10^{44}{\rm erg \, s^{-1}})$ \citep{Bentz2013}. The number density of BLR cloud is roughly in the range of $10^{8}{\rm cm^{-3}}$ and $10^{12}{\rm cm^{-3}}$\citep[e.g.,][]{Guo2020, Wu2023}, where the density can not be too low ($\gtrsim 10^7{\rm cm^{-3}}$) due to the absence of forbidden broad lines (e.g. [O {\footnotesize III}] $\lambda 4363$) or too high due to BLR will emit more continuum thermal emission than emission lines\citep[e.g.,][]{Korista2000}. In all Cloudy simulations, the grid steps in hydrogen density $(n)$ and distance to the continuum source $(r)$ are set to 0.2 dex. We take the abundance to be solar abundance and the cloud column density to be $N_{\rm H} = 10^{23}{\rm cm^{-2}}$ which is consistent with the general assumption in BLR photoionization simulations \citep{Dumont1998}.

\section{Result}\label{sec:result}

\begin{figure*}[ht]
\includegraphics[scale=0.7]{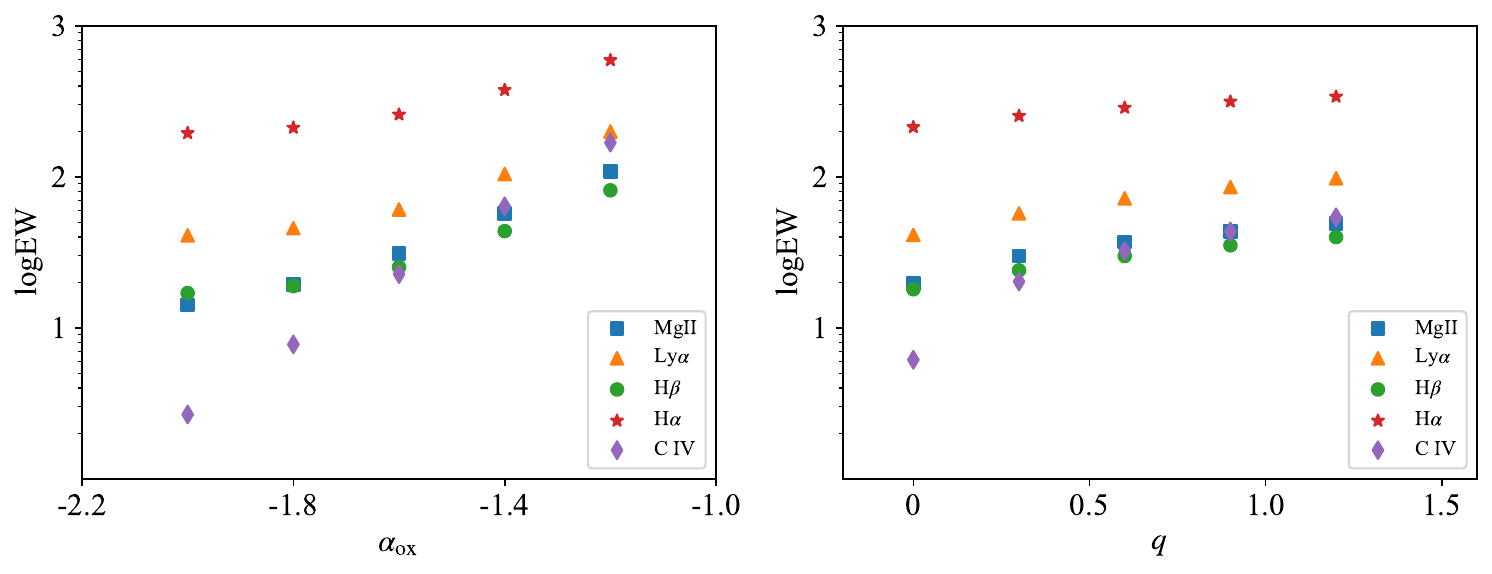}
\caption{The relations of EW--$\alpha_{\rm ox}$ (left panel) and EW--$q$(right panel) for several emission lines. The typical parameters ${\rm log}M_{\rm BH}/M_{\odot}=9$, $\dot{m}=0.1$, CF = 0.3, and $q=0.5$ (left panel) and $\alpha_{\rm ox}=-1.6$ (right panel) are adopted. 
\label{para_test}}
\end{figure*}

In Fig. \ref{para_test}, we present the relations of EW-$\alpha_{\rm ox}$ (left panel) and EW-$q$ (right panel) respectively for several typical emission lines (H$\alpha$, H$\beta$, Ly$\alpha$, Mg {\footnotesize II} and C {\footnotesize IV}). From the left panel, it can be found that the EW of all lines increases with increases of $\alpha_{\rm ox}$, where C {\footnotesize IV} lines are much more sensitive to the emission from soft to hard X-ray bands. For the case of varying $q$ from 0 to 1.2, the EW of H$\alpha$, H$\beta$, Ly$\alpha$ and Mg {\footnotesize II} increase less than 2.5 times, while EW of C {\footnotesize IV} increases more than 8 times (see right panel). The high ionization lines are more sensitive to these two parameters compared to the low ionization lines. Both low values of $\alpha_{\rm ox}$ and $q$, will lead to the weaker C {\footnotesize IV} lines (e.g., EW $<$ 10\text{\AA}). EW of Mg {\footnotesize II}, Ly$\alpha$ and H$\beta$ will also be less than $\sim 20-30 \text{\AA}$ for the case of steep optical to X-ray spectrum (e.g., $\alpha_{\rm ox}<-1.6$). Therefore, both high-ionization and low-ionization lines will become weaker with the decrease of soft X-ray intensity ($\alpha_{\rm ox}$ and $q$). It should be noted that the BH mass ${\rm log}M_{\rm BH}/M_{\odot}=9$ is adopted as an example, which is a typical value for WLQ (see Fig. \ref{para_test}). For lower BH mass of  ${\rm log}M_{\rm BH}/M_{\odot}=7-8$, this trend is roughly unchanged, however, all EWs are several tens larger (or no WLQs). 

\begin{figure*}[ht]
\includegraphics[scale=0.72]{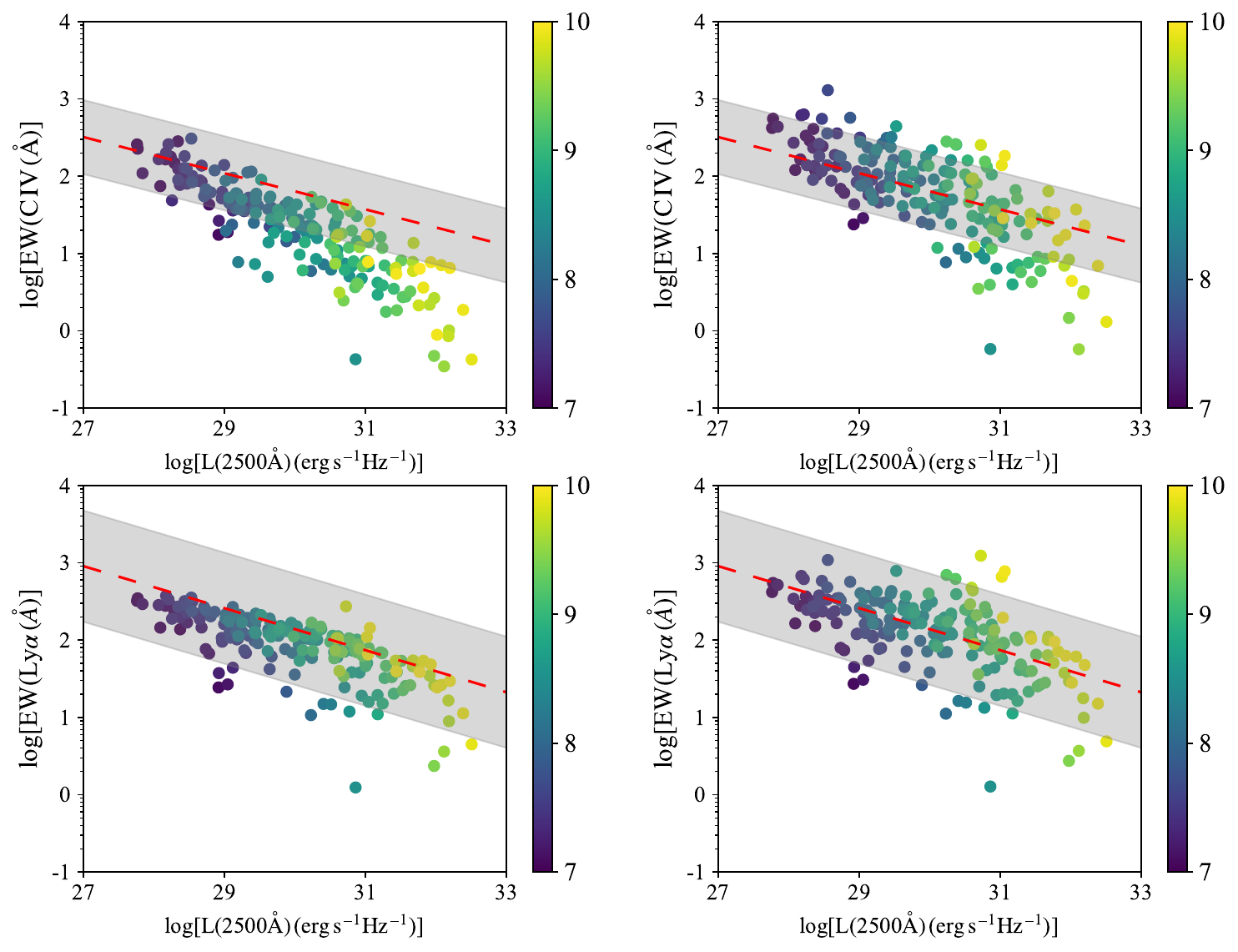}
\caption{The relation between $L(2500\text{\AA})$ and EW in our simulations for C {\footnotesize IV} (upper panels) and Ly$\alpha$ (lower panels). The left and right two panels represent the simulations without and with including the X-ray soft excess respectively. The dashed lines and grey region represent the observational results from \cite{Timlin2020} and \cite{Green2001} for EW(C {\footnotesize IV}) and EW(Ly$\alpha$), respectively. The colorbar is BH mass.
\label{l2500_CIV}}
\end{figure*}

In the left panel of Fig. \ref{l2500_CIV}, we present the results of 200 simulation points in $L(2500\text{\AA})$ - EW(C {\footnotesize IV}) relation without considering the soft X-ray excess (i.e., $q=0$). EW(C {\footnotesize IV}) follows an anti-correlation with $L(2500\text{\AA})$, which is consistent with the famous Baldwin effect. However, the EW(C {\footnotesize IV}) is systematically lower than those of observational results for normal quasars \citep[grey shades and red lines,][]{Timlin2020}. After including the soft X-ray component, the simulation results of $L(2500\text{\AA})$ - EW(C {\footnotesize IV}) correlation roughly consistent with that of the observational result (right panel in Fig. \ref{l2500_CIV}).  The slope of the simulation points in the right panel is -0.38 which is roughly consistent but slightly steeper than -0.23 from the observed $L(2500\text{\AA})$ - EW(C {\footnotesize IV}) relation. The sources with larger BH masses normally have smaller EW(C {\footnotesize IV}) and, therefore, the weak lines are more likely to occur in AGNs with heavier black holes (e.g., $10^9-10^{10}M_{\odot}$).

\begin{figure*}[ht]
\centering
\includegraphics[scale=0.72]{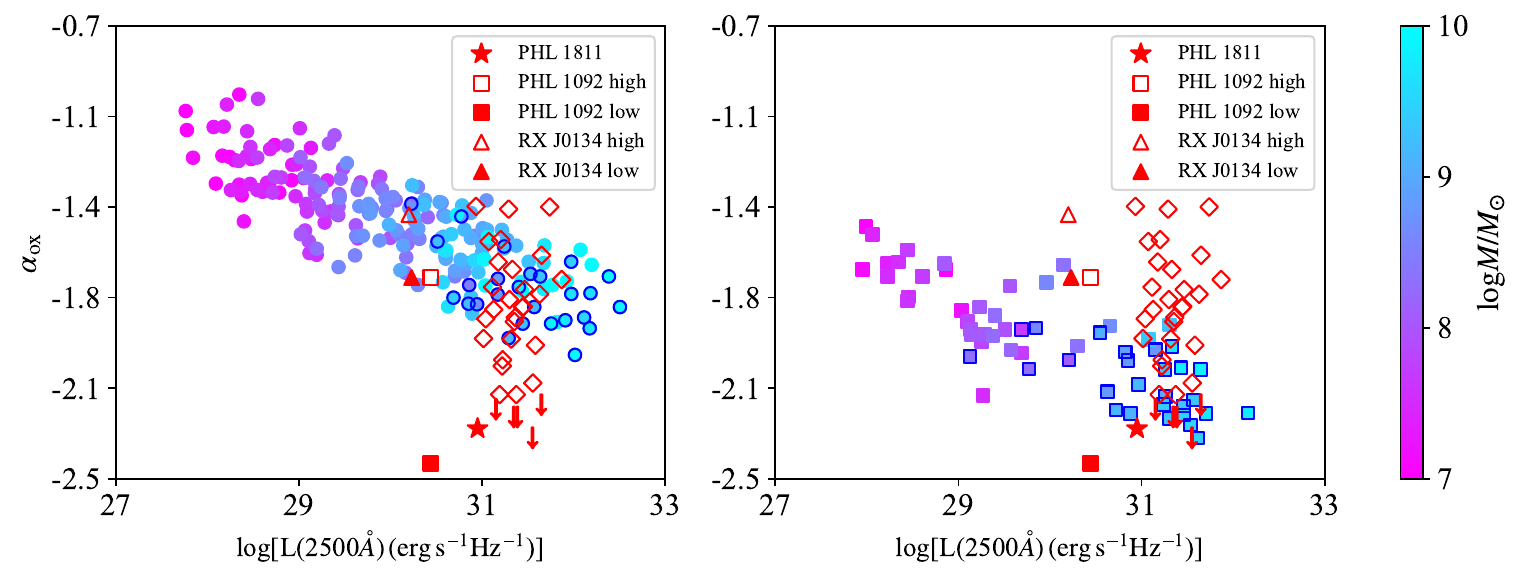}
\caption{The weak-line sources in $\alpha_{\rm ox}-L(2500\text{\AA})$ plane. In the left panel, the solid circles show our simulation result and the solid circles surrounded by hollow circles represent the weak-line AGNs in our simulations. For comparison, the observational data of some weak-line AGNs are also presented with the red hollow diamonds, which are selected from \cite{Ni2018, Ni2022}. In right panel, we also present 60 simulations with extremely weak X-ray emissions (solid squares), where $\alpha_{\rm ox}$ is smaller by 0.5 compared to that from relation $\alpha_{\rm ox}-L(2500\text{\AA})$, where solid squares surrounded by hollow squares also represent the weak-line AGNs in simulations.
\label{alpha_l2500}}
\end{figure*}

In Fig. \ref{alpha_l2500}, the circles represent the simulation results in $L(2500\text{\AA})$ - $\alpha_{\rm ox}$ diagram, where the solid circle enclosed by the blue circle represents that the equivalent width of C {\footnotesize IV} is less than $10\text{\AA}$ (i.e. WLQ). It can be found that the WLQs mostly stay in the lower right corner. The red diamond and downward arrow represent the observation WLQs sample and WLQs with X-ray upper limits from \cite{Ni2018, Ni2022}, where some WLQs have much lower $\alpha_{\rm ox}$ compared to our simulation parameter space. Therefore, we further simulate the line properties for $\alpha_{\rm ox}$ smaller than typical values by 0.5. We find that most of the simulated data belong to WLQs for $M_{\rm BH}>10^{8.5}M_{\odot}$, and there is also one WLQ with $M_{\rm BH}<10^8M_{\odot}$. This result suggests that AGNs with $M_{\rm BH}\sim10^8M_{\odot}$ can also show weak high-ionization lines when their soft/hard X-ray emission is extremely weak.

\begin{figure*}[htb]
\includegraphics[scale=0.50]{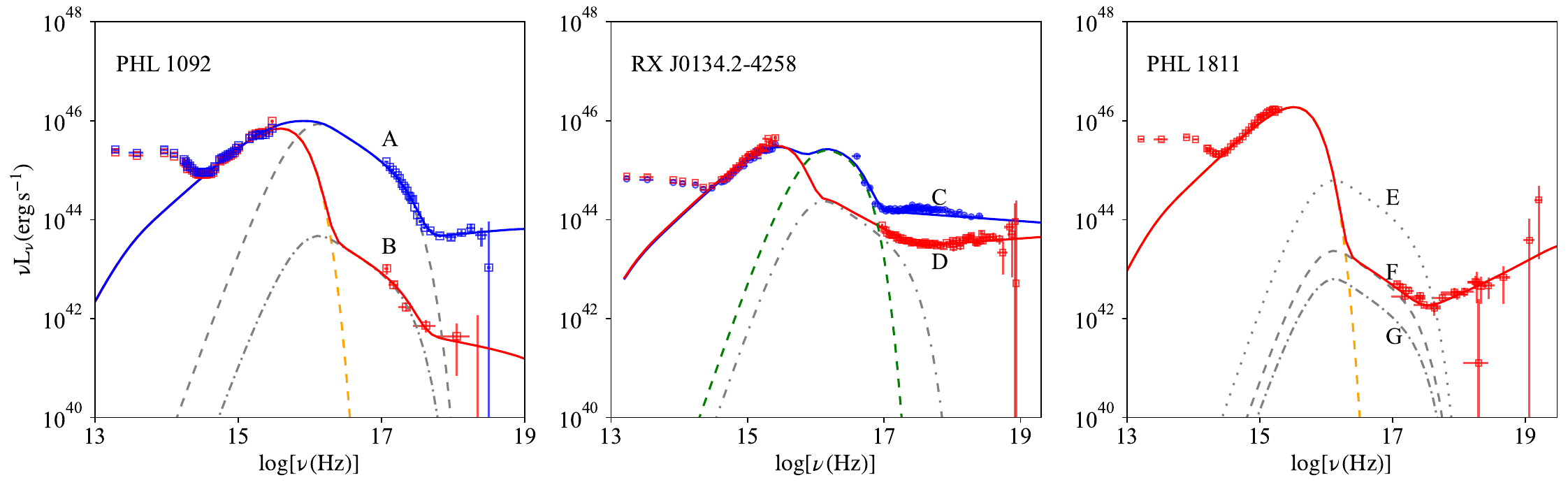}
\caption{The SED modeling for three weak-line AGNs, where the optical-UV emission is fitted with slim disk model while the X-ray emission is fitted by a soft X-ray component and a pow-law component. The high state and low state in PHL 1092 and RX J0134.2-4258 are fitted respectively. To test the model, we also consider two different strengths of soft X-ray components in PHL 1811.
\label{singleobj}}
\end{figure*}

To further test our model prediction, we also compare our model results with three WLQs as an example, where the multi-waveband observations and, in particular, the soft excess emission are detected. PHL 1811 and PHL 1092 are both well-known classic WLQs. RX J0134.2-4258 is currently the only known example of a Seyfert with similar weak UV lines. These three sources have excellent multi-band observational data available to reconstruct their SEDs, providing valuable inputs for our Cloudy simulations. We fit the spectral energy distribution (SED) for three weak-line AGNs and then explore the properties of their broad emission lines, where the multiband SEDs are adopted from \cite{Jin2023}. The continua are fitted with the above slim disk model, a soft X-ray component, and a power-law hard X-ray component. The fitting results and model parameters are presented in Fig. \ref{singleobj} and Table \ref{table1} respectively, where the SEDs of high state and low state are fitted respectively for PHL 1092 and RX J0134.2-4258 (hereafter, RX J0134). For PHL 1811, we also test the different strengths of soft X-ray excess since that its observed X-ray is very weak. In modeling the emission lines, we firstly constrain the covering factor with the strength of H$\beta$ lines \citep{Ferland2020}, and then give the predicted value of EW(C {\footnotesize IV}), where the results are presented in Table \ref{table1}. For PHL 1092 and RX J0134, the predicted values of EW(C {\footnotesize IV}) are 36.5 and 24.5 respectively, which do not belong to weak-line AGNs. However, the values of EW(C {\footnotesize IV}) are 12.1 and 7.4 in their low states, which are roughly consistent with the observed values. For PHL 1811, the predicted value of EW(C {\footnotesize IV}) is 4.8, which may stay in a low state since its X-ray emission is very weak. A slightly higher value of $q$ or stronger soft X-ray emission will lead to a larger value of EW(C {\footnotesize IV}).

\begin{deluxetable*}{lcccccccc}[ht]
\centering
\tabcolsep=0.2cm
\tablecaption{ Fitting parameters for a single object. \label{table1}}
\tablehead{
\colhead{Object Name/SED} & \colhead{Observed REW(H$\beta$)/\text{\AA}} & \colhead{Observed REW(C {\footnotesize IV})/\text{\AA}} & \colhead{CF} & \colhead{$M_{\rm BH}/M_{\odot}$} & \colhead{$\dot{m}$} & \colhead{$q$} & \colhead{$\alpha_{\rm ox}$} & \colhead{EW(C {\footnotesize IV})/\text{\AA}}
}
\startdata
PHL 1092/A & $26.1\pm5.4$ & $12.1\pm2.5$ & 0.1 & $3\times10^8$ & 0.6 & 10.7 & -1.7 & 36.5\\
PHL 1092/B & ... & ... & 0.2 & ... & ... & 4.2 & -2.4 & 12.1\\
RX J0134/C & $23.8\pm1.2$ & $7.6\pm1.5$ & 0.1 & $4\times10^8$ & 0.2 & $0.8^a$ & -1.5 & 24.5\\
RX J0134/D & ... & ... & 0.2 & ... & ... & 0.4 & -1.7 & 7.4\\
PHL 1811/E & $36.1\pm3.9$ & $7.9\pm2.5$ & 0.3 & $7\times10^8$ & 0.7 & 20.0 & -2.3 & 8.1\\
PHL 1811/F & ... & ... & 0.3 & ... & ... & 0.8 & ... & 4.8\\
PHL 1811/G & ... & ... & 0.3 & ... & ... & 0.2 & ... & 4.7\\
\enddata
\tablecomments{The rest frame equivalent width (REW) of observed lines. The last column is the EW(C {\footnotesize IV}) predicted by our model. All of the observational data come from \cite{Jin2023}. The BH mass $M_{\rm BH}$ and $\dot{m}$ are determined by SED fitting. $^a$The value $q$ of RX J0134.2-4258 in high state (C) is calculated based on 0.2-1 keV due to the the absence of soft excess above 1 keV. 
}
\end{deluxetable*}

\section{Conclusion and Discussion}\label{sec:discussion}

To understand the possible physical mechanism for the weak lines in a fraction of quasars, we investigate the properties of emission lines based on Cloudy code with AGN SEDs built from several observational empirical correlations, where we particularly consider the possible soft X-ray excess component. We find that the EW of high excitation lines (e.g., C {\footnotesize IV}, Ly$\alpha$, etc.) is normally smaller than the observational results if just considering the optical-UV emission from cold disk and power-law X-ray emission from corona. However, the theoretical predictions on EW(C {\footnotesize IV}) and EW(Ly$\alpha$) are consistent with the observations very well after considering the soft X-ray excess component as constrained from observations, which suggests that the soft X-ray emission is necessary in explaining the strength of high excitation lines. Several other parameters will also lead to weaker high excitation lines, which include steeper $\alpha_{\rm ox}$ (weaker hard/soft X-ray emission), unusual supermassive BHs (e.g., $M_{\rm BH} \gtrsim 10^9 M_{\odot}$) and/or low covering factor of BLR. 


It has been found that many AGNs show strong variability in the soft X-ray bands \citep{Miniutti2009, Jin2022, Jin2023}, even though their optical-UV emission is roughly unchanged (e.g., PHL 1092 and RX J0134 in this work). The variation in UV and soft X-ray wavebands will affect the ultraviolet high-ionization lines based on the photonionization model. Firstly, we explore the properties of the emission lines with different X-ray emissions that may stay in low or high states. We find that the C {\footnotesize IV} emission lines are strongly correlated with the strength of soft X-ray emission, where the weaker soft X-ray emission normally predicts much lower strength of C {\footnotesize IV} lines (see Table \ref{table1}). Both PHL 1811 and the low X-ray state of PHL 1092 and RX J0134 have a very small value of $\alpha_{\rm ox}$ (see Fig. \ref{alpha_l2500}), which suggests that their soft X-ray emission is very weak. Half of weak-line AGNs shown lower X-ray emission support this scenario. It should be noted that the observations of the  C {\footnotesize IV} lines and X-ray emission are not simultaneous in most WLQ sources. Future quasi-simultaneous observations on the optical spectrum and X-ray emission are wished to further test this issue, where the time delay between X-ray and emission line should be considered.

The power-law X-ray emission is widely believed to be produced via Comptonization of optical/UV seed photons from a standard accretion disk in an optically thin hot corona, where the intensity of power-law emission and power-law slope are determined by parameters like BH mass and accretion rate \citep{Cao2009, Netzer2013}. It is found that the soft X-ray emission below 2 keV is much higher than that extrapolated from the hard X-ray power-law emission, which is very common in type I AGN \citep{Pravdo1981, Singh1985, Walter1993, Piconcelli2005, Gliozzi2020}. The nature of the soft X-ray excess is poorly understood, where the most promising explanations include the warm corona scenario \citep[e.g.,][]{Chris2012} and relativistically smeared reflection scenario \citep[see e.g.,][]{Ballantyne2002, Gierlinski2004, Crummy2006}. In this work, we adopt a phenomenological model to mimic the soft X-ray emission in building the AGN SEDs. The physical origin of soft X-ray excess may differ from our model, but this will not affect our main conclusion. The soft excess may be very common in AGNs as suggested in \cite{Boissay2016}. The strength of the soft excess is positively correlated with the strength of the Eddington ratio, which is significantly larger in the narrow-line Seyfert I sources compared to the broad-line Seyfert I sources \citep[e.g.,][]{Boissay2016, Gliozzi2020}. The shape of the AGN SEDs plays a central role in determining both the ionization states and electron excitation in BLR clouds. In particular, the strength of UV lines strongly depends on the photons from UV to X-ray wavebands. \cite{Leighly2007_2} studied the emission lines in a WLQ of PHL 1811 by calculating the various SEDs and proposed that the weak collisionally excited high ionization line (e.g., C {\footnotesize IV}) can be attributed to the absence of EUV photons. In Fig. \ref{l2500_CIV}, it can be found that EW of UV lines of both Ly$\alpha$ and C {\footnotesize IV} in our simulations are statistically lower than the observational results if just considering the optical/UV component from the cold accretion disk and power-law component from the corona. However, the simulation results are roughly consistent with the observations very well after we include the component of the soft X-ray excess. It should be noted that the strength of the soft X-ray excess in WLQs is considered to be correlated with Eddington ratios as that constrained from local type I AGNs. More X-ray observations on the local WLQs and other observational constraints are expected to further test this issue. The adopted empirical correlations in the disk-corona model are found from normal AGNs, and it is still unclear whether WLQs have peculiar SED or not. For optical-UV emission, WLQs can well be described by standard disk model \citep{Marculewicz2020} and the relation between the hard X-ray index $\Gamma$ and accretion rate $\dot{m}$ still follow the relationship from normal AGNs \citep{Shemmer2010}.

Besides the soft X-ray excess component, several other factors may also lead to the weak lines in AGNs. The first one is the BH mass. The maximum effective temperature of the accretion disk is correlated with both BH mass and accretion rate. The exceptionally massive BHs will lead to lower disk temperature and a softer ionizing spectrum. \cite{Laor2011} proposed that the high-excitation line should be weak or even absent for non-spinning very massive BH AGNs, where the low-ionization lines may still be present. In Fig. \ref{l2500_CIV}, it can be found that most sources with EW(C {\footnotesize IV}) and EW(Ly$\alpha$) less than 10 normally have BH mass in range of $10^9-10^{10}M_{\odot}$ (right panels), which is consistent with results of \cite{Laor2011}. We note that the BH mass estimations in WLQs are still controversial, since their broad lines may be peculiar. Based on the continuum fitting method, \cite{Marculewicz2020} found that most of the WLQs have BH mass larger than $10^9M_{\odot}$ and pointed out that the BH mass estimated from FWHM(H$\beta$) is underestimated. However, \cite{Ha2023} proposed that the BH mass estimated from the traditional method of broad-emission-line region size–luminosity relation may be overestimated up to an order of magnitude, where the corrections based on the strength of the optical Fe {\footnotesize II} emission and possible blue-shift of C {\footnotesize IV} lines are applied. It should be noted that the larger value of  $R_{\rm Fe\, {\footnotesize II}}= $Fe\,{\footnotesize II}(4434-4684\text{\AA})/H$\beta$ in WLQs may not be an indicator of high Eddington ratios as found in narrow-line Seyferts, but caused by the weaker H$\beta$ emission since both low-ionization and high-ionization lines can become weaker in WLQs (see Fig. \ref{para_test} and \cite{Shemmer2010}). Therefore, further investigations on the BH mass of WLQs will help understand their intrinsic physics.


The second parameter is $\alpha_{\rm ox}$, which controls the strength of X-ray emission. It is well known that $\alpha_{\rm ox}$ is anti-correlated with the UV luminosity (e.g., $L(2500\text{\AA})$ - $\alpha_{\rm ox}$), which may be triggered by the stronger corona cooling for sources with higher disk luminosity. If this is the case, it suggests that high BH mass and high Eddington ratios will lead to strong UV luminosity and weak X-ray emission. From Fig. \ref{para_test}, it can be found that the steeper $\alpha_{\rm ox}$ will lead to smaller EW, where we have assumed the soft X-ray emission is a certain fraction of hard X-ray emission (i.e., a fixed $q$ parameter). It was found that half of the WLQs show weaker X-ray emission, where they deviate from the $L(2500\text{\AA})$ - $\alpha_{\rm ox}$ relation \citep[e.g.,][]{Luo2015, Ni2018, Ni2022, Paul2022}. The special weak X-ray emission in these WLQs is not so clear, which can be intrinsic or obscured by the shielding gas (e.g., torus-like slim disk or optically thick winds). We further do a couple of simulations with weaker X-ray emission, where $\alpha_{\rm ox}$ is reduced by 0.5. We find that roughly all simulations with this extremely weak X-ray emission are weak line AGNs for $M_{\rm BH}>10^9M_{\odot}$, where one source with $M_{\rm BH}<10^8M_{\odot}$ also belong to weak line AGNs. It should be noted that the strength of soft X-ray emission is controlled by the parameters of both $\alpha_{\rm ox}$ and $q$. The soft X-ray emission may still be weak if parameter $q$, which is determined by the correlation with large intrinsic scatter, is small at given $\alpha_{\rm ox}$, which may be the possible physical reason for some WLQs still follow $L(2500\text{\AA})$ - $\alpha_{\rm ox}$ relation.

The third parameter is the covering factor of BLR, which will affect the line intensities. \cite{Nikolajuk2012} proposed that the covering factor of WLQs may be around one order of magnitude lower than that of normal AGNs based on the ratios of high-ionization line and low-ionization line regions. \cite{Kumar2023} found that the BLR in WLQs is underdeveloped, where the covering factor of WLQs is about two times lower than that of normal AGNs based on the assumption that the covering factors of the BLR and the dusty torus have to be the same \citep[e.g.,][]{Gaskell2007}. The fraction of infrared emission is used to infer the torus covering factors in WLQs, which is reasonable if the emission from the other band is the same. But the UV to soft X-ray emission of WLQs is lower than that of normal quasars\citep[see Figure 2 in][]{Kumar2023}, which roughly supports our conclusion for weak UV-soft X-ray emissions in WLQs. Therefore, the lower infrared emission in WLQs may be not fully caused by the low covering factor of the torus, where their intrinsic bolometric luminosity may be low. Furthermore, the difference of two times on the covering factor is still not enough to explain the weak lines in most WLQs. In this work, we adopt the cloud covering factor decrease as the Eddington ratio increases, where the covering factor is around 80\% at low-luminosity AGNs and decreases to several percent at Eddington ratios as that constrained from the H$\beta$ and He {\footnotesize II} lines. Therefore, the fraction of WLQs is quite high in our simulations, which is mainly caused by the low covering factor for $\dot{M}>\dot{M}_{\rm Edd}$. Better constraints on the covering factor from observations for wide parameter space of BH mass and Eddington ratio are wished to further explore this issue.


We explore the UV line properties in AGNs based on Cloudy simulations after considering AGN SEDs with wide parameter space of BH mass, accretion rate, and soft X-ray excess, where the UV to soft X-ray spectrum is important for these UV lines. Most of the works assumed that one or two parameters may dominate the physical reasons for the weak lines. Based on our simulation, a large fraction of quasars with BH mass $M_{\rm BH}>10^9M_{\odot}$ might be WLQs if do not consider the possible contribution from the soft X-ray excess. Based on observations, the quasars with very massive BHs (e.g., $M_{\rm BH}>10^9M_{\odot}$) normally have weaker extreme UV emission and weaker soft X-ray emission (or steeper $\alpha_{\rm ox}$), which normally cannot excite strong UV lines. Therefore, we propose that the intensity of soft X-ray excess plays an important role in the intensity of UV lines. The AGNs with very heavier BHs, or extreme X-ray emission in both soft and hard X-ray wavebands, or very low covering factor of BLR can also lead to weak UV lines. The broad-band observations on AGN SED and emission lines (both high ionization lines and low ionization lines)  can help to distinguish the above properties. We adopt several observed correlations from normal quasars in exploring the line properties, where we assume that most properties of WLQs may be similar to those of normal quasars. We have tested several possible parameters (even not all), where more massive BH mass, lower accretion rate, weaker soft X-ray emission (steeper $\alpha_{\rm ox}$ or weaker hard X-ray) will possibly lead to weaker high-ionization lines. It is possible that WLQs may stay in a special stage of AGNs, where the empirical correlations as constrained from the normal quasars cannot applied to WLQs. Most of the WLQs are found in high redshift (z$>$2) objects, which may be an artifact of the SDSS optical spectral survey. \cite{Paul2022} explored the WLQs in intermediate redshift (0.9 $<$ z $<$ 1.5) based on the ultraviolet spectroscopy of HST, where the weak high ionization lines are more or less similar to those of high-redshift WLQs \citep[e.g. see also][]{Leighly2007, Wu2012, Plotkin2010_2}. UV spectroscopy surveys can find more nearby WLQs, which will help to explore the possible evolution of WLQs. Though hundreds of WLQs have been found in SDSS, the WLQs sample may be still incomplete due to observation bias. In our model, we just investigate the possible affection of AGN SEDs on the line intensities, where we cannot rule out the possibility that WLQs may stay in a special stage of AGNs (e.g., BLR is not well-formed).

\begin{acknowledgments}
The work is supported by the National Natural Science Foundation of China (grants 12233007 and U1931203) and the science research grants from the China Manned Space Project (No. CMS-CSST-2021-A06). The authors acknowledge Beijng PARATERA Tech CO., Ltd. for providing HPC resources that have contributed to the results reported within this paper.
\end{acknowledgments}

%

\vspace{5mm}


\software{Cloudy \citep{Chatzikos2023}
          }

\bibliography{sample631}{}

\begin{thebibliography}{}
\expandafter\ifx\csname natexlab\endcsname\relax\def\natexlab#1{#1}\fi
\providecommand{\url}[1]{\href{#1}{#1}}
\providecommand{\dodoi}[1]{doi:~\href{http://doi.org/#1}{\nolinkurl{#1}}}
\providecommand{\doeprint}[1]{\href{http://ascl.net/#1}{\nolinkurl{http://ascl.net/#1}}}
\providecommand{\doarXiv}[1]{\href{https://arxiv.org/abs/#1}{\nolinkurl{https://arxiv.org/abs/#1}}}

\bibitem[{{Abramowicz} {et~al.}(1988){Abramowicz}, {Czerny}, {Lasota}, \& {Szuszkiewicz}}]{Abramowicz1988}
{Abramowicz}, M.~A., {Czerny}, B., {Lasota}, J.~P., \& {Szuszkiewicz}, E. 1988, \apj, 332, 646, \dodoi{10.1086/166683}

\bibitem[{{Anderson} {et~al.}(2001){Anderson}, {Fan}, {Richards}, {Schneider}, {Strauss}, {Vanden Berk}, {Gunn}, {Knapp}, {Schlegel}, {Voges}, {Yanny}, {Bahcall}, {Bernardi}, {Brinkmann}, {Brunner}, {Csab{\'a}i}, {Doi}, {Fukugita}, {Hennessy}, {Ivezi{\'c}}, {Kunszt}, {Lamb}, {Loveday}, {Lupton}, {McKay}, {Munn}, {Nichol}, {Szokoly}, \& {York}}]{Anderson2001}
{Anderson}, S.~F., {Fan}, X., {Richards}, G.~T., {et~al.} 2001, \aj, 122, 503, \dodoi{10.1086/321168}

\bibitem[{{Antonucci}(1993)}]{Antonucci1993}
{Antonucci}, R. 1993, \araa, 31, 473, \dodoi{10.1146/annurev.aa.31.090193.002353}

\bibitem[{{Baldwin} {et~al.}(1995){Baldwin}, {Ferland}, {Korista}, \& {Verner}}]{Baldwin1995}
{Baldwin}, J., {Ferland}, G., {Korista}, K., \& {Verner}, D. 1995, \apjl, 455, L119, \dodoi{10.1086/309827}

\bibitem[{{Ballantyne} {et~al.}(2002){Ballantyne}, {Ross}, \& {Fabian}}]{Ballantyne2002}
{Ballantyne}, D.~R., {Ross}, R.~R., \& {Fabian}, A.~C. 2002, in X-ray Spectroscopy of AGN with Chandra and XMM-Newton, ed. T.~{Boller}, S.~{Komossa}, S.~{Kahn}, H.~{Kunieda}, \& L.~{Gallo}, 73, \dodoi{10.48550/arXiv.astro-ph/0204260}

\bibitem[{{Bentz} {et~al.}(2013){Bentz}, {Denney}, {Grier}, {Barth}, {Peterson}, {Vestergaard}, {Bennert}, {Canalizo}, {De Rosa}, {Filippenko}, {Gates}, {Greene}, {Li}, {Malkan}, {Pogge}, {Stern}, {Treu}, \& {Woo}}]{Bentz2013}
{Bentz}, M.~C., {Denney}, K.~D., {Grier}, C.~J., {et~al.} 2013, \apj, 767, 149, \dodoi{10.1088/0004-637X/767/2/149}

\bibitem[{{Boissay} {et~al.}(2016){Boissay}, {Ricci}, \& {Paltani}}]{Boissay2016}
{Boissay}, R., {Ricci}, C., \& {Paltani}, S. 2016, \aap, 588, A70, \dodoi{10.1051/0004-6361/201526982}

\bibitem[{{Bottorff} {et~al.}(2002){Bottorff}, {Baldwin}, {Ferland}, {Ferguson}, \& {Korista}}]{Bottorff2002}
{Bottorff}, M.~C., {Baldwin}, J.~A., {Ferland}, G.~J., {Ferguson}, J.~W., \& {Korista}, K.~T. 2002, \apj, 581, 932, \dodoi{10.1086/344408}

\bibitem[{{Brightman} {et~al.}(2013){Brightman}, {Silverman}, {Mainieri}, {Ueda}, {Schramm}, {Matsuoka}, {Nagao}, {Steinhardt}, {Kartaltepe}, {Sanders}, {Treister}, {Shemmer}, {Brandt}, {Brusa}, {Comastri}, {Ho}, {Lanzuisi}, {Lusso}, {Nandra}, {Salvato}, {Zamorani}, {Akiyama}, {Alexander}, {Bongiorno}, {Capak}, {Civano}, {Del Moro}, {Doi}, {Elvis}, {Hasinger}, {Laird}, {Masters}, {Mignoli}, {Ohta}, {Schawinski}, \& {Taniguchi}}]{Brightman2013}
{Brightman}, M., {Silverman}, J.~D., {Mainieri}, V., {et~al.} 2013, \mnras, 433, 2485, \dodoi{10.1093/mnras/stt920}

\bibitem[{{Cao}(2009)}]{Cao2009}
{Cao}, X. 2009, \mnras, 394, 207, \dodoi{10.1111/j.1365-2966.2008.14347.x}

\bibitem[{{Chand} {et~al.}(2014){Chand}, {Kumar}, \& {Gopal-Krishna}}]{Chand2014}
{Chand}, H., {Kumar}, P., \& {Gopal-Krishna}. 2014, \mnras, 441, 726, \dodoi{10.1093/mnras/stu624}

\bibitem[{{Chatzikos} {et~al.}(2023){Chatzikos}, {Bianchi}, {Camilloni}, {Chakraborty}, {Gunasekera}, {Guzm{\'a}n}, {Milby}, {Sarkar}, {Shaw}, {van Hoof}, \& {Ferland}}]{Chatzikos2023}
{Chatzikos}, M., {Bianchi}, S., {Camilloni}, F., {et~al.} 2023, arXiv e-prints, arXiv:2308.06396, \dodoi{10.48550/arXiv.2308.06396}

\bibitem[{{Collinge} {et~al.}(2005){Collinge}, {Strauss}, {Hall}, {Ivezi{\'c}}, {Munn}, {Schlegel}, {Zakamska}, {Anderson}, {Harris}, {Richards}, {Schneider}, {Voges}, {York}, {Margon}, \& {Brinkmann}}]{Collinge2005}
{Collinge}, M.~J., {Strauss}, M.~A., {Hall}, P.~B., {et~al.} 2005, \aj, 129, 2542, \dodoi{10.1086/430216}

\bibitem[{{Crummy} {et~al.}(2006){Crummy}, {Fabian}, {Gallo}, \& {Ross}}]{Crummy2006}
{Crummy}, J., {Fabian}, A.~C., {Gallo}, L., \& {Ross}, R.~R. 2006, \mnras, 365, 1067, \dodoi{10.1111/j.1365-2966.2005.09844.x}

\bibitem[{{Czerny}(2019)}]{Czerny2019}
{Czerny}, B. 2019, Universe, 5, 131, \dodoi{10.3390/universe5050131}

\bibitem[{{Diamond-Stanic} {et~al.}(2009){Diamond-Stanic}, {Fan}, {Brandt}, {Shemmer}, {Strauss}, {Anderson}, {Carilli}, {Gibson}, {Jiang}, {Kim}, {Richards}, {Schmidt}, {Schneider}, {Shen}, {Smith}, {Vestergaard}, \& {Young}}]{Aleksandar2009}
{Diamond-Stanic}, A.~M., {Fan}, X., {Brandt}, W.~N., {et~al.} 2009, \apj, 699, 782, \dodoi{10.1088/0004-637X/699/1/782}

\bibitem[{{Done} {et~al.}(2012){Done}, {Davis}, {Jin}, {Blaes}, \& {Ward}}]{Chris2012}
{Done}, C., {Davis}, S.~W., {Jin}, C., {Blaes}, O., \& {Ward}, M. 2012, \mnras, 420, 1848, \dodoi{10.1111/j.1365-2966.2011.19779.x}

\bibitem[{{Dumont} {et~al.}(1998){Dumont}, {Collin-Souffrin}, \& {Nazarova}}]{Dumont1998}
{Dumont}, A.-M., {Collin-Souffrin}, S., \& {Nazarova}, L. 1998, \aap, 331, 11

\bibitem[{{Ezhikode} {et~al.}(2017){Ezhikode}, {Gandhi}, {Done}, {Ward}, {Dewangan}, {Misra}, \& {Philip}}]{Ezhikode2017}
{Ezhikode}, S.~H., {Gandhi}, P., {Done}, C., {et~al.} 2017, \mnras, 472, 3492, \dodoi{10.1093/mnras/stx2160}

\bibitem[{{Fan} {et~al.}(1999){Fan}, {Strauss}, {Gunn}, {Lupton}, {Carilli}, {Rupen}, {Schmidt}, {Moustakas}, {Davis}, {Annis}, {Bahcall}, {Brinkmann}, {Brunner}, {Csabai}, {Doi}, {Fukugita}, {Heckman}, {Hennessy}, {Hindsley}, {Ivezi{\'c} }, {Knapp}, {Lamb}, {Munn}, {Pauls}, {Pier}, {Rockosi}, {Schneider}, {Szalay}, {Tucker}, \& {York}}]{Fan1999}
{Fan}, X., {Strauss}, M.~A., {Gunn}, J.~E., {et~al.} 1999, \apjl, 526, L57, \dodoi{10.1086/312382}

\bibitem[{{Feng} {et~al.}(2019){Feng}, {Cao}, {Gu}, \& {Ma}}]{Feng2019}
{Feng}, J., {Cao}, X., {Gu}, W.-M., \& {Ma}, R.-Y. 2019, \apj, 885, 93, \dodoi{10.3847/1538-4357/ab4592}

\bibitem[{{Ferguson} {et~al.}(1997){Ferguson}, {Korista}, {Baldwin}, \& {Ferland}}]{Ferguson1997}
{Ferguson}, J.~W., {Korista}, K.~T., {Baldwin}, J.~A., \& {Ferland}, G.~J. 1997, \apj, 487, 122, \dodoi{10.1086/304611}

\bibitem[{{Ferland}(2003)}]{Ferland2003}
{Ferland}, G.~J. 2003, \araa, 41, 517, \dodoi{10.1146/annurev.astro.41.011802.094836}

\bibitem[{{Ferland} {et~al.}(2020){Ferland}, {Done}, {Jin}, {Landt}, \& {Ward}}]{Ferland2020}
{Ferland}, G.~J., {Done}, C., {Jin}, C., {Landt}, H., \& {Ward}, M.~J. 2020, \mnras, 494, 5917, \dodoi{10.1093/mnras/staa1207}

\bibitem[{{Ferland} {et~al.}(1998){Ferland}, {Korista}, {Verner}, {Ferguson}, {Kingdon}, \& {Verner}}]{Ferland1998}
{Ferland}, G.~J., {Korista}, K.~T., {Verner}, D.~A., {et~al.} 1998, \pasp, 110, 761, \dodoi{10.1086/316190}

\bibitem[{{Gaskell} {et~al.}(2007){Gaskell}, {Klimek}, \& {Nazarova}}]{Gaskell2007}
{Gaskell}, C.~M., {Klimek}, E.~S., \& {Nazarova}, L.~S. 2007, arXiv e-prints, arXiv:0711.1025, \dodoi{10.48550/arXiv.0711.1025}

\bibitem[{{Gierli{\'n}ski} \& {Done}(2004)}]{Gierlinski2004}
{Gierli{\'n}ski}, M., \& {Done}, C. 2004, \mnras, 349, L7, \dodoi{10.1111/j.1365-2966.2004.07687.x}

\bibitem[{{Giustini} \& {Proga}(2019)}]{Giustini2019}
{Giustini}, M., \& {Proga}, D. 2019, \aap, 630, A94, \dodoi{10.1051/0004-6361/201833810}

\bibitem[{{Gliozzi} \& {Williams}(2020)}]{Gliozzi2020}
{Gliozzi}, M., \& {Williams}, J.~K. 2020, \mnras, 491, 532, \dodoi{10.1093/mnras/stz3005}

\bibitem[{{Gopal-Krishna} {et~al.}(2013){Gopal-Krishna}, {Joshi}, \& {Chand}}]{Gopal2013}
{Gopal-Krishna}, {Joshi}, R., \& {Chand}, H. 2013, \mnras, 430, 1302, \dodoi{10.1093/mnras/sts706}

\bibitem[{{Green} {et~al.}(2001){Green}, {Forster}, \& {Kuraszkiewicz}}]{Green2001}
{Green}, P.~J., {Forster}, K., \& {Kuraszkiewicz}, J. 2001, \apj, 556, 727, \dodoi{10.1086/321600}

\bibitem[{{Guo} {et~al.}(2020){Guo}, {Shen}, {He}, {Wang}, {Liu}, {Wang}, {Sun}, {Yang}, {Kong}, \& {Sheng}}]{Guo2020}
{Guo}, H., {Shen}, Y., {He}, Z., {et~al.} 2020, \apj, 888, 58, \dodoi{10.3847/1538-4357/ab5db0}

\bibitem[{{Ha} {et~al.}(2023){Ha}, {Dix}, {Matthews}, {Shemmer}, {Brotherton}, {Myers}, {Richards}, {Maithil}, {Anderson}, {Brandt}, {Diamond-Stanic}, {Fan}, {Gallagher}, {Green}, {Lira}, {Luo}, {Netzer}, {Plotkin}, {Runnoe}, {Schneider}, {Strauss}, {Trakhtenbrot}, \& {Wu}}]{Ha2023}
{Ha}, T., {Dix}, C., {Matthews}, B.~M., {et~al.} 2023, \apj, 950, 97, \dodoi{10.3847/1538-4357/acd04d}

\bibitem[{{Hryniewicz} {et~al.}(2010){Hryniewicz}, {Czerny}, {Niko{\l}ajuk}, \& {Kuraszkiewicz}}]{Hryniewicz2010}
{Hryniewicz}, K., {Czerny}, B., {Niko{\l}ajuk}, M., \& {Kuraszkiewicz}, J. 2010, \mnras, 404, 2028, \dodoi{10.1111/j.1365-2966.2010.16418.x}

\bibitem[{{Jiang} {et~al.}(2019){Jiang}, {Blaes}, {Stone}, \& {Davis}}]{Jiang2019}
{Jiang}, Y.-F., {Blaes}, O., {Stone}, J.~M., \& {Davis}, S.~W. 2019, \apj, 885, 144, \dodoi{10.3847/1538-4357/ab4a00}

\bibitem[{{Jin} {et~al.}(2022){Jin}, {Done}, {Ward}, {Panessa}, {Liu}, \& {Liu}}]{Jin2022}
{Jin}, C., {Done}, C., {Ward}, M., {et~al.} 2022, \mnras, 512, 5642, \dodoi{10.1093/mnras/stac827}

\bibitem[{{Jin} {et~al.}(2023){Jin}, {Done}, {Ward}, {Panessa}, {Liu}, \& {Liu}}]{Jin2023}
---. 2023, \mnras, 518, 6065, \dodoi{10.1093/mnras/stac3513}

\bibitem[{{Korista} \& {Goad}(2000)}]{Korista2000}
{Korista}, K.~T., \& {Goad}, M.~R. 2000, \apj, 536, 284, \dodoi{10.1086/308930}

\bibitem[{{Kumar} {et~al.}(2016){Kumar}, {Chand}, \& {Gopal-Krishna}}]{Kumar2016}
{Kumar}, P., {Chand}, H., \& {Gopal-Krishna}. 2016, \mnras, 461, 666, \dodoi{10.1093/mnras/stw1374}

\bibitem[{{Kumar} \& {Gopal-Krishna}(2015)}]{Kumar2015}
{Kumar}, P., \& {Gopal-Krishna}, Chand, H. 2015, \mnras, 448, 1463, \dodoi{10.1093/mnras/stv060}

\bibitem[{{Kumar} {et~al.}(2023){Kumar}, {Chand}, \& {Joshi}}]{Kumar2023}
{Kumar}, R., {Chand}, H., \& {Joshi}, R. 2023, \mnras, 519, 3656, \dodoi{10.1093/mnras/stac3689}

\bibitem[{{Lane} {et~al.}(2011){Lane}, {Shemmer}, {Diamond-Stanic}, {Fan}, {Anderson}, {Brandt}, {Plotkin}, {Richards}, {Schneider}, \& {Strauss}}]{Lane2011}
{Lane}, R.~A., {Shemmer}, O., {Diamond-Stanic}, A.~M., {et~al.} 2011, \apj, 743, 163, \dodoi{10.1088/0004-637X/743/2/163}

\bibitem[{{Laor} \& {Davis}(2011)}]{Laor2011}
{Laor}, A., \& {Davis}, S.~W. 2011, \mnras, 417, 681, \dodoi{10.1111/j.1365-2966.2011.19310.x}

\bibitem[{{Leighly} {et~al.}(2007{\natexlab{a}}){Leighly}, {Halpern}, {Jenkins}, \& {Casebeer}}]{Leighly2007_2}
{Leighly}, K.~M., {Halpern}, J.~P., {Jenkins}, E.~B., \& {Casebeer}, D. 2007{\natexlab{a}}, \apjs, 173, 1, \dodoi{10.1086/519768}

\bibitem[{{Leighly} {et~al.}(2007{\natexlab{b}}){Leighly}, {Halpern}, {Jenkins}, {Grupe}, {Choi}, \& {Prescott}}]{Leighly2007}
{Leighly}, K.~M., {Halpern}, J.~P., {Jenkins}, E.~B., {et~al.} 2007{\natexlab{b}}, \apj, 663, 103, \dodoi{10.1086/518017}

\bibitem[{{Liu} {et~al.}(2013){Liu}, {Zhang}, \& {Zhang}}]{Liu2013}
{Liu}, Y., {Zhang}, J., \& {Zhang}, S.-N. 2013, in Feeding Compact Objects: Accretion on All Scales, ed. C.~M. {Zhang}, T.~{Belloni}, M.~{M{\'e}ndez}, \& S.~N. {Zhang}, Vol. 290, 267--268, \dodoi{10.1017/S1743921312019941}

\bibitem[{{Londish} {et~al.}(2004){Londish}, {Heidt}, {Boyle}, {Croom}, \& {Kedziora-Chudczer}}]{Londish2004}
{Londish}, D., {Heidt}, J., {Boyle}, B.~J., {Croom}, S.~M., \& {Kedziora-Chudczer}, L. 2004, \mnras, 352, 903, \dodoi{10.1111/j.1365-2966.2004.07980.x}

\bibitem[{{Luo} {et~al.}(2015){Luo}, {Brandt}, {Hall}, {Wu}, {Anderson}, {Garmire}, {Gibson}, {Plotkin}, {Richards}, {Schneider}, {Shemmer}, \& {Shen}}]{Luo2015}
{Luo}, B., {Brandt}, W.~N., {Hall}, P.~B., {et~al.} 2015, \apj, 805, 122, \dodoi{10.1088/0004-637X/805/2/122}

\bibitem[{{Marculewicz} \& {Nikolajuk}(2020)}]{Marculewicz2020}
{Marculewicz}, M., \& {Nikolajuk}, M. 2020, \apj, 897, 117, \dodoi{10.3847/1538-4357/ab9597}

\bibitem[{{McDowell} {et~al.}(1995){McDowell}, {Canizares}, {Elvis}, {Lawrence}, {Markoff}, {Mathur}, \& {Wilkes}}]{McDowell1995}
{McDowell}, J.~C., {Canizares}, C., {Elvis}, M., {et~al.} 1995, \apj, 450, 585, \dodoi{10.1086/176168}

\bibitem[{{Miniutti} {et~al.}(2009){Miniutti}, {Fabian}, {Brandt}, {Gallo}, \& {Boller}}]{Miniutti2009}
{Miniutti}, G., {Fabian}, A.~C., {Brandt}, W.~N., {Gallo}, L.~C., \& {Boller}, T. 2009, \mnras, 396, L85, \dodoi{10.1111/j.1745-3933.2009.00669.x}

\bibitem[{{Murray} {et~al.}(1995){Murray}, {Chiang}, {Grossman}, \& {Voit}}]{Murray1995}
{Murray}, N., {Chiang}, J., {Grossman}, S.~A., \& {Voit}, G.~M. 1995, \apj, 451, 498, \dodoi{10.1086/176238}

\bibitem[{{Netzer}(2013)}]{Netzer2013}
{Netzer}, H. 2013, {The Physics and Evolution of Active Galactic Nuclei}

\bibitem[{{Ni} {et~al.}(2018){Ni}, {Brandt}, {Luo}, {Hall}, {Shen}, {Anderson}, {Plotkin}, {Richards}, {Schneider}, {Shemmer}, \& {Wu}}]{Ni2018}
{Ni}, Q., {Brandt}, W.~N., {Luo}, B., {et~al.} 2018, \mnras, 480, 5184, \dodoi{10.1093/mnras/sty1989}

\bibitem[{{Ni} {et~al.}(2022){Ni}, {Brandt}, {Luo}, {Garmire}, {Hall}, {Plotkin}, {Shemmer}, {Timlin}, {Vito}, {Wu}, \& {Yi}}]{Ni2022}
---. 2022, \mnras, 511, 5251, \dodoi{10.1093/mnras/stac394}

\bibitem[{{Niko{\l}ajuk} \& {Walter}(2012)}]{Nikolajuk2012}
{Niko{\l}ajuk}, M., \& {Walter}, R. 2012, \mnras, 420, 2518, \dodoi{10.1111/j.1365-2966.2011.20216.x}

\bibitem[{{Paul} {et~al.}(2022){Paul}, {Plotkin}, {Shemmer}, {Anderson}, {Brandt}, {Fan}, {Gallo}, {Luo}, {Ni}, {Richards}, {Schneider}, {Wu}, \& {Yi}}]{Paul2022}
{Paul}, J.~D., {Plotkin}, R.~M., {Shemmer}, O., {et~al.} 2022, \apj, 929, 78, \dodoi{10.3847/1538-4357/ac5bd6}

\bibitem[{{Piconcelli} {et~al.}(2005){Piconcelli}, {Jimenez-Bail{\'o}n}, {Guainazzi}, {Schartel}, {Rodr{\'\i}guez-Pascual}, \& {Santos-Lle{\'o}}}]{Piconcelli2005}
{Piconcelli}, E., {Jimenez-Bail{\'o}n}, E., {Guainazzi}, M., {et~al.} 2005, \aap, 432, 15, \dodoi{10.1051/0004-6361:20041621}

\bibitem[{{Plotkin} {et~al.}(2010{\natexlab{a}}){Plotkin}, {Anderson}, {Brandt}, {Diamond-Stanic}, {Fan}, {MacLeod}, {Schneider}, \& {Shemmer}}]{Plotkin2010_2}
{Plotkin}, R.~M., {Anderson}, S.~F., {Brandt}, W.~N., {et~al.} 2010{\natexlab{a}}, \apj, 721, 562, \dodoi{10.1088/0004-637X/721/1/562}

\bibitem[{{Plotkin} {et~al.}(2010{\natexlab{b}}){Plotkin}, {Anderson}, {Brandt}, {Diamond-Stanic}, {Fan}, {Hall}, {Kimball}, {Richmond}, {Schneider}, {Shemmer}, {Voges}, {York}, {Bahcall}, {Snedden}, {Bizyaev}, {Brewington}, {Malanushenko}, {Malanushenko}, {Oravetz}, {Pan}, \& {Simmons}}]{Plotkin2010}
---. 2010{\natexlab{b}}, \aj, 139, 390, \dodoi{10.1088/0004-6256/139/2/390}

\bibitem[{{Plotkin} {et~al.}(2015){Plotkin}, {Shemmer}, {Trakhtenbrot}, {Anderson}, {Brandt}, {Fan}, {Gallo}, {Lira}, {Luo}, {Richards}, {Schneider}, {Strauss}, \& {Wu}}]{Plotkin2015}
{Plotkin}, R.~M., {Shemmer}, O., {Trakhtenbrot}, B., {et~al.} 2015, \apj, 805, 123, \dodoi{10.1088/0004-637X/805/2/123}

\bibitem[{{Pravdo} {et~al.}(1981){Pravdo}, {Nugent}, {Nousek}, {Jensen}, {Wilson}, \& {Becker}}]{Pravdo1981}
{Pravdo}, S.~H., {Nugent}, J.~J., {Nousek}, J.~A., {et~al.} 1981, \apj, 251, 501, \dodoi{10.1086/159489}

\bibitem[{{Shakura} \& {Sunyaev}(1973)}]{Shakura1973}
{Shakura}, N.~I., \& {Sunyaev}, R.~A. 1973, \aap, 24, 337

\bibitem[{{Shemmer} {et~al.}(2006){Shemmer}, {Brandt}, {Schneider}, {Fan}, {Strauss}, {Diamond-Stanic}, {Richards}, {Anderson}, {Gunn}, \& {Brinkmann}}]{Shemmer2006}
{Shemmer}, O., {Brandt}, W.~N., {Schneider}, D.~P., {et~al.} 2006, \apj, 644, 86, \dodoi{10.1086/503543}

\bibitem[{{Shemmer} {et~al.}(2010){Shemmer}, {Trakhtenbrot}, {Anderson}, {Brandt}, {Diamond-Stanic}, {Fan}, {Lira}, {Netzer}, {Plotkin}, {Richards}, {Schneider}, \& {Strauss}}]{Shemmer2010}
{Shemmer}, O., {Trakhtenbrot}, B., {Anderson}, S.~F., {et~al.} 2010, \apjl, 722, L152, \dodoi{10.1088/2041-8205/722/2/L152}

\bibitem[{{Singh} {et~al.}(1985){Singh}, {Garmire}, \& {Nousek}}]{Singh1985}
{Singh}, K.~P., {Garmire}, G.~P., \& {Nousek}, J. 1985, \apj, 297, 633, \dodoi{10.1086/163560}

\bibitem[{{Steffen} {et~al.}(2006){Steffen}, {Strateva}, {Brandt}, {Alexander}, {Koekemoer}, {Lehmer}, {Schneider}, \& {Vignali}}]{Steffen2006}
{Steffen}, A.~T., {Strateva}, I., {Brandt}, W.~N., {et~al.} 2006, \aj, 131, 2826, \dodoi{10.1086/503627}

\bibitem[{{Timlin} {et~al.}(2020){Timlin}, {Brandt}, {Ni}, {Luo}, {Pu}, {Schneider}, {Vivek}, \& {Yi}}]{Timlin2020}
{Timlin}, J.~D., {Brandt}, W.~N., {Ni}, Q., {et~al.} 2020, \mnras, 492, 719, \dodoi{10.1093/mnras/stz3433}

\bibitem[{{Titarchuk}(1994)}]{Titarchuk1994}
{Titarchuk}, L. 1994, \apj, 434, 570, \dodoi{10.1086/174760}

\bibitem[{{Urry} \& {Padovani}(1995)}]{Urry1995}
{Urry}, C.~M., \& {Padovani}, P. 1995, \pasp, 107, 803, \dodoi{10.1086/133630}

\bibitem[{{Walter} \& {Fink}(1993)}]{Walter1993}
{Walter}, R., \& {Fink}, H.~H. 1993, \aap, 274, 105

\bibitem[{{Wu} {et~al.}(2012){Wu}, {Brandt}, {Anderson}, {Diamond-Stanic}, {Hall}, {Plotkin}, {Schneider}, \& {Shemmer}}]{Wu2012}
{Wu}, J., {Brandt}, W.~N., {Anderson}, S.~F., {et~al.} 2012, \apj, 747, 10, \dodoi{10.1088/0004-637X/747/1/10}

\bibitem[{{Wu} {et~al.}(2023){Wu}, {Wu}, {Xue}, {Lei}, \& {Lyu}}]{Wu2023}
{Wu}, J., {Wu}, Q., {Xue}, H., {Lei}, W., \& {Lyu}, B. 2023, \apj, 950, 106, \dodoi{10.3847/1538-4357/acce9e}

\bibitem[{{Wu} {et~al.}(2011){Wu}, {Brandt}, {Hall}, {Gibson}, {Richards}, {Schneider}, {Shemmer}, {Just}, \& {Schmidt}}]{Wu2011}
{Wu}, J., {Brandt}, W.~N., {Hall}, P.~B., {et~al.} 2011, \apj, 736, 28, \dodoi{10.1088/0004-637X/736/1/28}

\bibitem[{{Zhang} \& {Liu}(2016)}]{Zhang2016}
{Zhang}, X., \& {Liu}, Y. 2016, \apj, 830, 69, \dodoi{10.3847/0004-637X/830/2/69}

\end{thebibliography}
\bibliographystyle{aasjournal}

\end{document}